 \documentclass[final,3p,times,twocolumn]{elsarticle}

\usepackage{amssymb}
\usepackage{amsmath}
\usepackage{color}

\journal{}

\begin{document}

\begin{frontmatter}



\title{Ultra-Thin SiGe in the Source Modifies Performance of Thin-Film Tunneling FET 
}

\author{Hamed Habibi}
\author{Shoeib Babaee Touski}
\ead{touski@hut.ac.ir}
\address{Department of Electrical Engineering, Hamedan University of Technology, Hamedan 65155, Iran}

\begin{abstract}
In this work, the source structure of an n-type thin-film tunneling FET is engineered to get better performance. An ultra-thin SiGe along with Si is used in the source of silicon-based TFET. Two structures are compared with conventional TFET, one, SiGe is located on the top of Si in the source and another one in reverse. Simulations approve these structures can reduce sub-threshold swing, OFF-current several times, and increase the ON-OFF ratio. Band diagram for conduction and valance bands are investigated and band to band tunneling (BTBT) generation rate is used to find better performance. We find current flows at Si in the source with the wider bandgap. Ge mole fraction of SiGe is varied and its effects on the performance of TFET are studied. The SiGe thickness for both structures is explored to obtain the best thickness for SiGe. 
\end{abstract}


\end{frontmatter}


\section{Introduction}
In the past decades, conventional MOSFET has been scaled down to obtain low-power, high-speed operations, and compactness of electronics equipment. However, shrinkage of conventional MOSFET is limited by short channel effect, variation in silicon thickness, random dopant fluctuations, and high sub-threshold swing ($SS$) \cite{taur98,chaudhry04,leung12}. These issues cause to drain-induced barrier lowering, low $\mathrm{I_{ON}}$/$\mathrm{I_{OFF}}$ ratio, high power dissipation due to $\mathrm{I_{OFF}}$ increasing and especially high sub-threshold swing \cite{bangsaruntip10}. Sub-threshold swing is limited to $\mathrm{60mV/decade}$ for conventional MOS transistors. The thermionic injection of electrons over the energy barrier which limits the transition slope from the OFF to the ON-state. To achieve a low OFF-state current with a low threshold voltage, the sub-threshold slope should be smaller than $\mathrm{60mV/decade}$. however, this is not possible with the MOSFET.

High sub-threshold swing is a major concern for MOSFET technology. To overcome this issue, various possible devices have been explored in which tunnel FETs (TFETs) have attracted huge attention \cite{wang04}. TFETs are proposed for future generation low-power devices due to their low subthreshold swing ($SS$) and low OFF-current \cite{choi07,ionescu11}. TFET is suggested the best for low standby power family in the 2005 ITRS \cite{ITRS,esseni14}.

TFET suffers from some challenges, including low ON current ($\mathrm{I_{ON}}$) and high ambipolar conduction \cite{wu2015short}. The low $\mathrm{I_{ON}}$ hinders application in high-speed and RF applications. The low $\mathrm{I_{ON}}$ value in Si-TFETs is owing to a weak BTBT rate due to a large and indirect bandgap, and high effective mass of the carriers \cite{bagga15}. The low $\mathrm{I_{ON}}$ issue has been settled by various strategies such as the use of lower band-gap materials \cite{mookerjea08}, III-V TFETs with staggered/broken bandgaps \cite{zhou2012novel,koswatta10}, high-mobility channel material in strained Ge \cite{krishnamohan08}, and source doping optimizations \cite{toh08device}.
Several gate and dielectric engineering have also been proposed like dual-metal gate TFETs \cite{vishnoi14,wang2010design}, double-gate TFETs with a tri-material gate [11], gate-all-around triple-metal TFETs \cite{bagga17}, hetero-gate dielectric TFETs \cite{choi2010hetero}, and dual-material gate hetero-dielectric TFETs.

With decreasing tunneling barrier to increasing the on-current, the smaller tunneling barrier also leads to higher OFF-current 
\cite{brocard13,villalon14}. Furthermore, $\mathrm{I_{OFF}}$, as well as $SS$, would be further degraded by the enhanced trap-assisted tunneling (TAT) owing to the large density of interface defects in HTFETs \cite{schenk17}.
The consequent degradation of the OFF-current in heterojunction can also be alleviated by the double metal gate technique \cite{toh08device,wang2010design}.

Strained Si has been investigated to increase $\mathrm{I_{ON}}$ \cite{nayfeh2008design} whereas this material is highly affected by the growth method. For enhancing the ON-current, recently indium arsenide (InAs) is proposed because of its low bandgap However it  suffers from OFF-current \cite{ahish2015performance,strangio2016}. Ge in the source with a low bandgap is another choice to improve ON-current. Ge is one of the promising candidate materials due to the high hole and electron mobility and it also is compatible with Si technology \cite{claeys2011germanium}. An epi-layer Ge on a Si substrate as the channel for TFET can reduce $SS$ and enhance its performance \cite{lee2013hetero}. Patel, et al \cite{patel2008drive}, proposed SiGe at the top of the source region can enhance the performance of TFET. They proposed the best thickness for SiGe is 40nm. Whereas, the thickness of the thin film transistor reaches to bellow this value. Chunlei Wu, et al \cite{wu2016novel} split drain to two regions, half Si and half Ge, whereas Ge region is located under the Si region far from the gate. They reduce the average sub-threshold swing with this hetero-structure TFET. All these works used Ge or SiGe in the source of n-type TFET to decline tunneling path in ON-state and increase ON-current. Here, we also use SiGe in the source of n-type thin film TFET and explore its effects on the performance.

In this work, three structures for n-type TFET with source material engineering are studied. The best structure with better performance is investigated. SiGe is used instead of Ge and the structures are investigated for different mole fraction. In the end, the best SiGe thickness is explored.

\begin{figure}[t]
	\centering
	\includegraphics[width=1.0\linewidth]{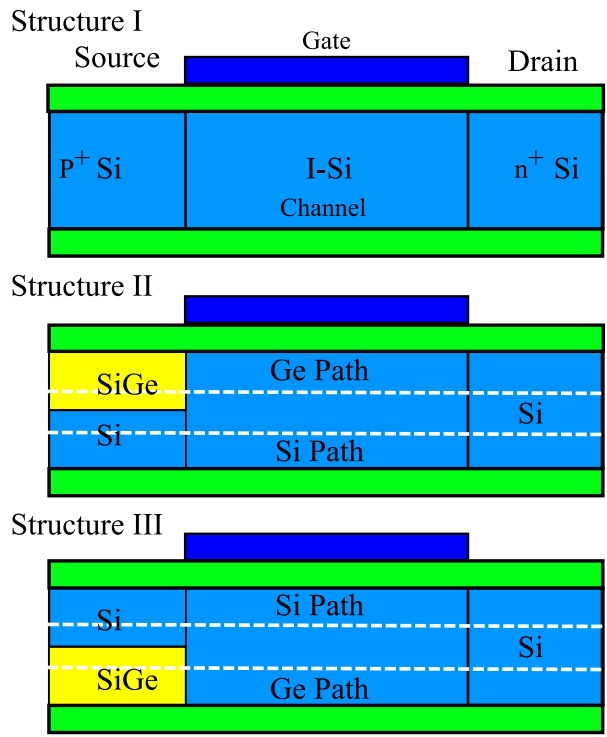}
	\caption{Schematic of three studied structures.}
	\label{fig:fig1}
\end{figure} 

\section{MODEL AND FORMALISM}
The proposed structures are investigated using a device simulator as implanted in Atlas Silvaco \cite{silvaco}. In our simulation, non-local band-to-band tunneling (BTBT) model \cite{esseni14,choi07}, is used that calculates the BTBT rate at each point. However, SRH and Auger models are used in the simulation to account carrier recombination effects. Band gap narrowing (BGN) model is activated for consideration the effects of high doping concentration on the reduction of the band gap. Electric field dependent Lombardi (CVT) model is turned on to consider mobility reduction in a high electric field. The simulations use a very fine mesh across the tunneling region. The proposed devices are simulated with Silvaco and current is obtained for different states.

\begin{table}[t]
	\caption{SiGe parameters is modeled with equations using experimental reports. \cite{busch60,fischetti96,ershov94,john69,canali75,ottaviani75,ryder53,liou96}}
	\centering
	\begin{tabular}{l l} 
		\hline\hline 
		\rule{0pt}{3ex} 
		Symbol~~~~~~~~~~~~ &  ~~~~~~~~~~Equation \rule{0pt}{2ex} \\
		
		\hline \rule{0pt}{3ex}
		$\mu_n$     & $\mu_n(x)=\exp(7.37-10.90x+11.51x^2)$ \\
		\rule{0pt}{2ex}
		$\mu_p$     & $\mu_p(x)=\exp(6.35-5.97x+6.97x^2)$ \\
		\rule{0pt}{2ex}
		$v_{sat,n}$ & $v_{sat,n}(x)=10^{7}/(0.98+2.93x-2.39x^2)$ \\ 
		\rule{0pt}{2ex}
		$v_{sat,p}$ & $v_{sat,p}(x)=10^{7}/(1.36+1.91x-1.88x^2)$  \\ 
		\hline\hline 
	\end{tabular}
	\label{table:tab1} 
\end{table}

\begin{table}[t]
	\caption{Energy band parameters for Si$_{1-x}$Ge$_x$ \cite{yu2018ultrathin}.}
	\centering
	\begin{tabular}{l l l l l l l} 
		\hline\hline 
		\rule{0pt}{3ex} 
		x & 0 (Si) & 0.2 & 0.4 & 0.6 & 0.8 & 1.0 (Ge) \\
		\hline 
		\rule{0pt}{3ex}
		$E_g[eV]$  & 1.1 & 1.0 & 0.93 & 0.89 & 0.86 & 0.66 \\
		$\chi$[eV] & 4.05 & 4.01 & 3.94 & 3.83 & 3.86 & 3.95 \\
		\hline\hline 
	\end{tabular}
	\label{table:tab2} 
\end{table}

The electrical properties of Si$_{1-x}$Ge$_x$ vary with Ge mole fraction (x) and these properties should be inserted in the simulations. Their properties change from Si ($x=0$) to Ge ($x=1$). Here, the modeling of mobility ($\mu$) and saturation velocity ($v_{sat}$) for strained SiGe on Si is taken from Ref. \cite{yu2018ultrathin}. Material parameters for SiGe are used from experimental results and the amount for $\mu$ and $v_{sat}$ are derived from a SiGe layer grown on a Si substrate. The mobility and velocity saturation equations as a function of Ge mole fraction (x) are listed in Table \ref{table:tab1}. The model for both electron and hole is included in the Table.

Band alignment should be considered in hetero-structure systems. Electron affinity ($\chi$) and bandgap ($E_g$) determines the energy of conduction and valance band. The location of bands in SiGe relative to Si controls band alignment and behavior of the heterostructure. The electron affinity and bandgap of Si$_{1-x}$Ge$_x$ for different $x$ values are listed in Table. \ref{table:tab2}. The parameters extracted from the experimental reported and inserted in the simulation. Band gap varies from $\mathrm{1.1eV}$ for $\mathrm{x=0}$ (Si) to $\mathrm{0.66eV}$ for $\mathrm{x=1}$ (Ge) and electron affinity also changes from $\mathrm{4.05eV}$ for Si to $\mathrm{3.95eV}$ for Ge.

\section{results and discussion}
Material engineering at the source gives a chance to enhance the performance of a TFET. We compare three different structures for TFET, see Fig. \ref{fig:fig1}. The first one is a common TFET that all source, channel, and drain are constructed with Silicon. The second structure, the source is split into two regions, up region close to top-gate is constructed from SiGe and underneath material from Si. The Third structure is inverse of the second structure so that silicon is used on the top of SiGe in the source. Source, drain, and channel lengths are considered 5nm, 5nm, and 10nm, respectively. Gate oxide thickness is considered 1nm and the gate is considered poly-Si. Doping concentration for source, channel and drain regions are considered $\mathrm{2\times 10^{20} cm^{-3}}$, $\mathrm{10^{17} cm^{-3}}$, and $\mathrm{2\times 10^{19} cm^{-3}}$, respectively. Drain-source voltage is considered as $\mathrm{V_{DS}=0.1V}$. $\mathrm{V_{GS}=-0.75V}$ and $\mathrm{V_{GS}=-3V}$ are selected for OFF- and ON-state, respectively.

\begin{figure}[t]
	\centering
	\includegraphics[width=0.8\linewidth]{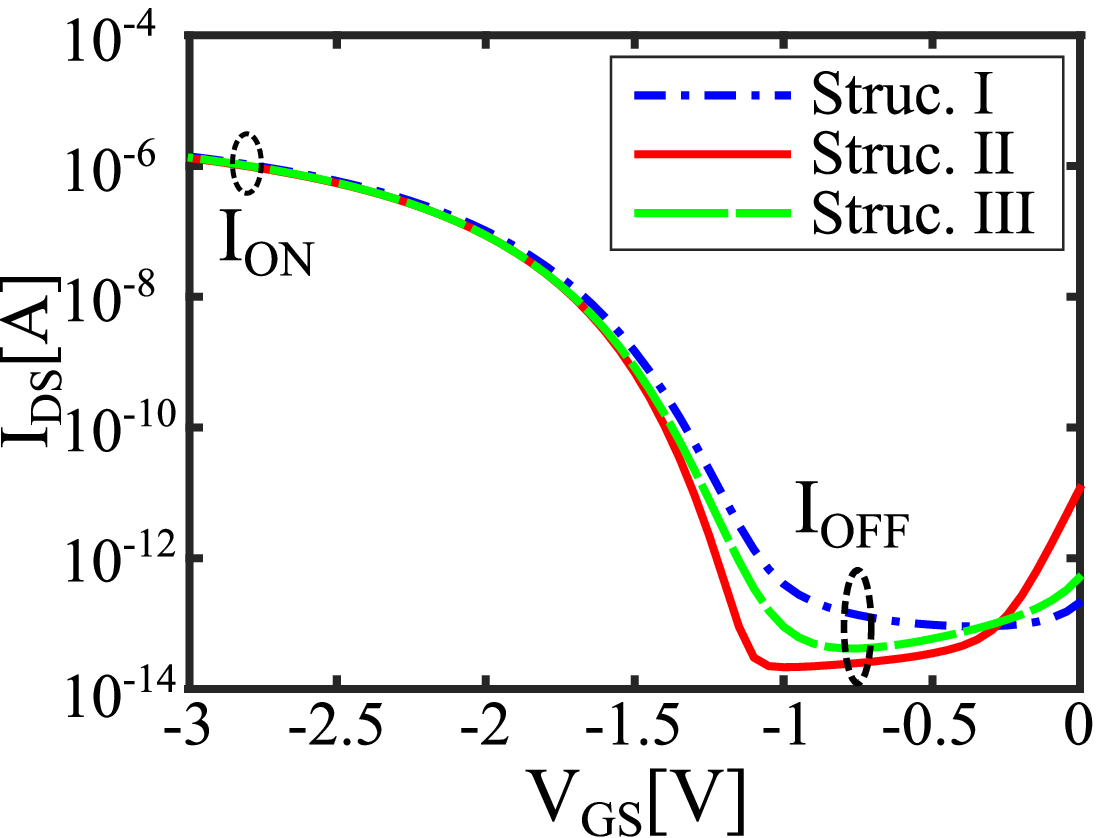}
	\caption{  Drain current versus gate voltage for three structures. $\mathrm{L_S=L_D=5nm}$,  $\mathrm{L_{Ch}=10nm}$.} 
	\label{fig:fig2}
\end{figure}

The drain current versus gate voltage for different structures is shown in Fig. \ref{fig:fig2}. Three structures are compared with the main parameters- $\mathrm{I_{ON}}$,  $\mathrm{I_{OFF}}$, ON-OFF ratio, and sub-threshold swing ($SS$). $SS$ in the sub-threshold regime can be defined as:
\begin{equation}
SS=\frac{V_{GS,2}-V_{GS,1}}{log(I_{DS,2})-log(I_{DS,1})}
\end{equation}
This parameter indicates the current slope when the transistor switches on. The sub-threshold swing, $\mathrm{I_{ON}}$, $\mathrm{I_{OFF}}$ and ON-OFF ratio for structures are reported in Table \ref{table:tab3}. ON-current for all structures is the same and close to each other. Structure II shows the lowest of $\mathrm{I_{ON}}$ and conventional TFET the highest. On the other hand, Structure II shows the lowest $\mathrm{I_{OFF}}$, it is four times lower than conventional TFET. Structure III also displays a low OFF-current, two times lower than conventional TEFT. Due to the lowest OFF-current in structure II, this structure shows the highest ON-OFF ratio. The ON-OFF ratio for structure II is four times larger than two other structures.

The main concern for TFET is $SS$ that small $SS$ can decline voltage and power supply. Structure III reduces $SS$ from $\mathrm{117mv/dec}$ for conventional TFET to $\mathrm{105mV/dec}$ that this reduction is not much. On the other hand, structure II decreases $SS$ from $\mathrm{117mV/dec}$ to $\mathrm{71mV/dec}$ which means this structure can decline $SS$ as 50$\%$. Structure III behaves highly close to conventional TFET. These results approve the structure II can be selected for future TFET. A FET with this structure is simulated to compare with the proposed structures. $\mathrm{I_{ON}}$, $\mathrm{I_{OFF}}$, ON-OFF ratio and sub-threshold swing are obtained as $2.61 mA$, $0.127\mu A$, $2.05\times 10^{4}$,  and $\mathrm{468mV/dec}$, respectively. With comparing with the proposed structures, ON-current is larger than the proposed structures, whereas, its OFF-current is approximately four times lower than the conventional TFET. This causes the ON-OFF ratio to rise more than three times. The sub-threshold swing declines from 117mV/dec in the conventional TFET to 71 in the proposed TFET.  These results indicate the performance of planar FET will halt for short channel length where TFET (our proposed structures) can continue planar processes.

\begin{table}[t]
	\caption{TFET properties for three structures. The unit of $SS$ is mV/dec.}
	\centering
	\begin{tabular}{l l l l l} 
		\hline\hline 
		\rule{0pt}{4ex} 
		&  $I_{ON}[A]$ & $I_{OFF}[A]$ & $I_{ON}/I_{OFF}$ & $SS$ \rule{0pt}{1ex}\vspace{0.4mm}  \\
		\hline
		\rule{0pt}{3ex} 
		Struc. I  &  1.41e-6 & 9.05e-14 & 1.56e7 & 117 \\
		Struc. II & 1.33e-6 & 2.49e-14 & 5.33e7 & 71 \\
		Struc. III  & 1.36e-6 & 4.16e-14 & 3.27e7 & 105 \\
		\hline\hline 
	\end{tabular}
	\label{table:tab3} 
\end{table}

\begin{figure*}
	\centering
	\includegraphics[width=1\linewidth]{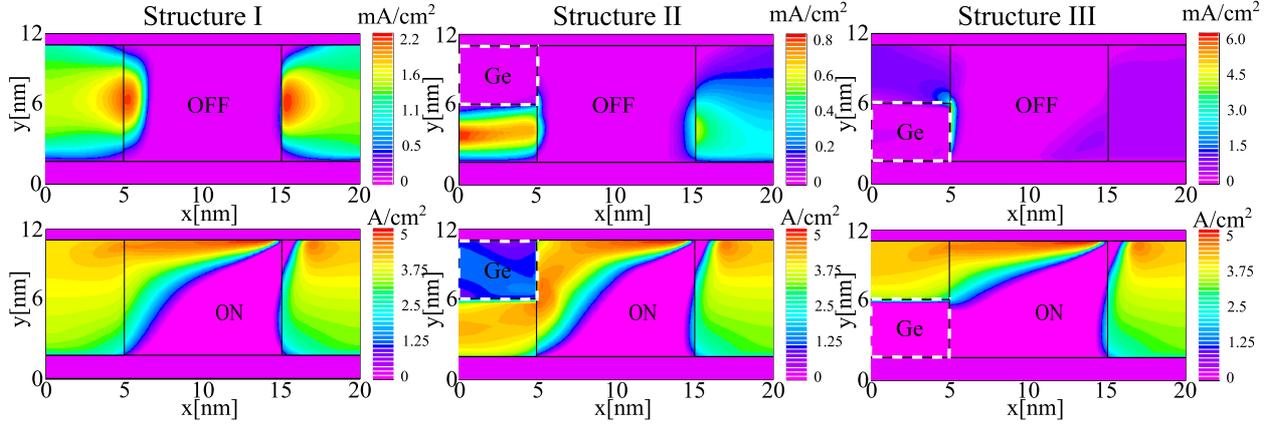}
	\caption{Current density in the structures at ON-state and OFF-state. The upper plots are for OFF-current and bottoms for ON-current. Ge regions are highlighted in the plots.}
	\label{fig:fig3}
\end{figure*}

For clarifying the performance of the proposed structures, map current density distribution for all structures are plotted in Fig. \ref{fig:fig3}. The OFF-state current for the structures is plotted in the top of the figure and ON-state in the bottom. The germanium region is highlighted in the figures for better following. As one can see, OFF-state current for conventional TFET (structure I) passes in the whole source region. On the contrary, ON-current mainly passes on the top of the channel near to the gate. One can expect that $\mathrm{I_{ON}}$ increases by using Ge on the top of source with higher mobility but we see from table \ref{table:tab3} that ON current approximately is equal for the structures. 

\begin{figure}
	\centering
	\includegraphics[width=1.0\linewidth]{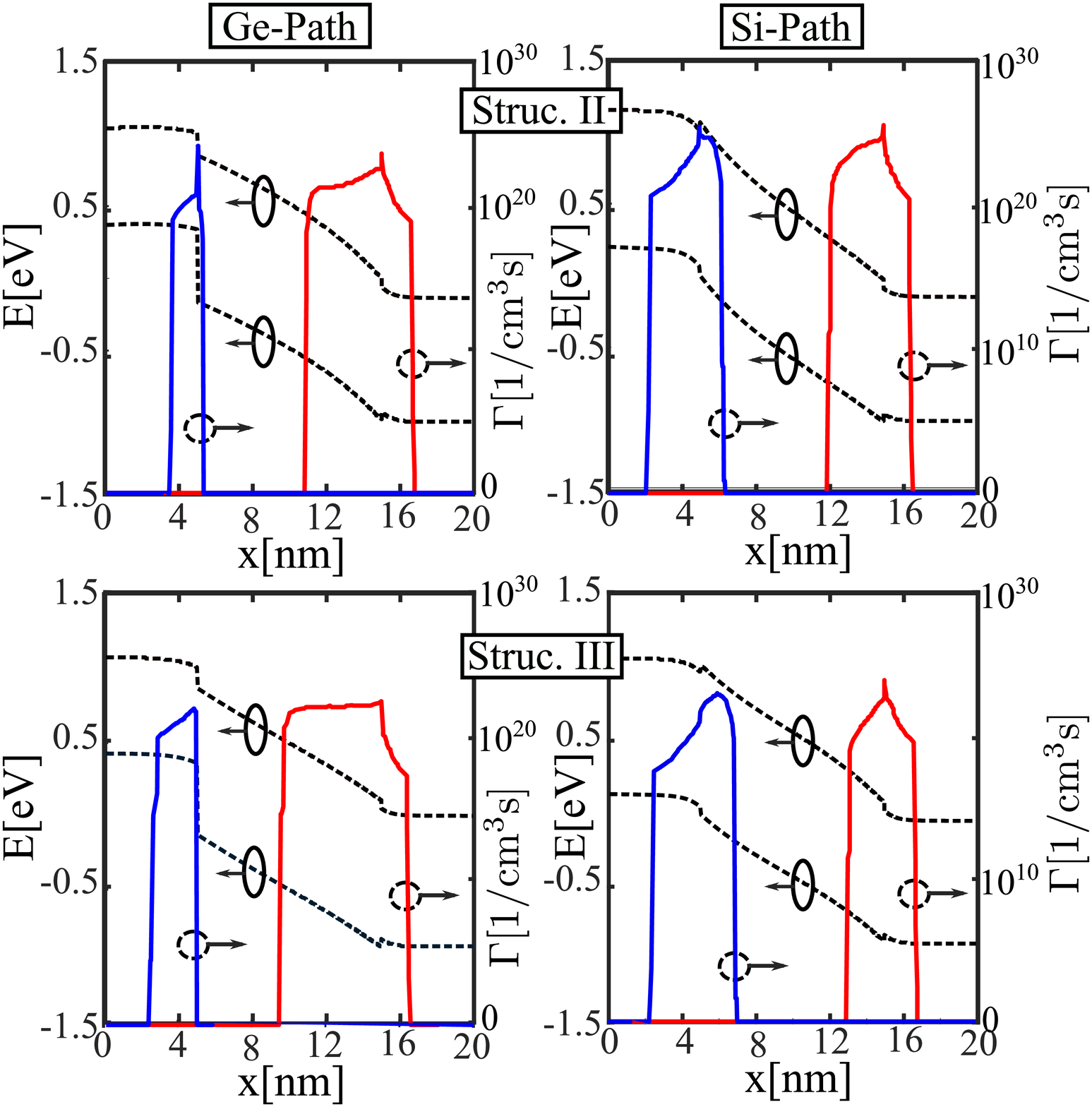}
	\caption{Energy band diagram and BTBT rate along the channel for OFF-current. Figures are for structure II  at the middle of (a) Ge and (b) Si regions in the source and for structure III at the middle of (c) Si and (d) Si.}
	\label{fig:fig4}
\end{figure}

One can observe in structure II, both OFF- and ON-current in the source highly pass through Si with a wider bandgap and lower mobility. We need further investigation to gain a better understanding of this behavior. In structure III, a high part of OFF-current flows in the Si region near to gate and a small part in the Ge region. ON-current mainly passes through the Si region. One can understand by comparing two structures that $\mathrm{I_{OFF}}$ highly likes to flow far from the gate and ON-current passes in the Si region. We expected that $\mathrm{I_{ON}}$ passes through the Ge region with lower bandgap but here ON-current passes in the Si region with a wider bandgap. Although, Ge is close to the gate and has a lower bandgap, current highly flows in the Si region.

\begin{figure}
	\centering
	\includegraphics[width=1\linewidth]{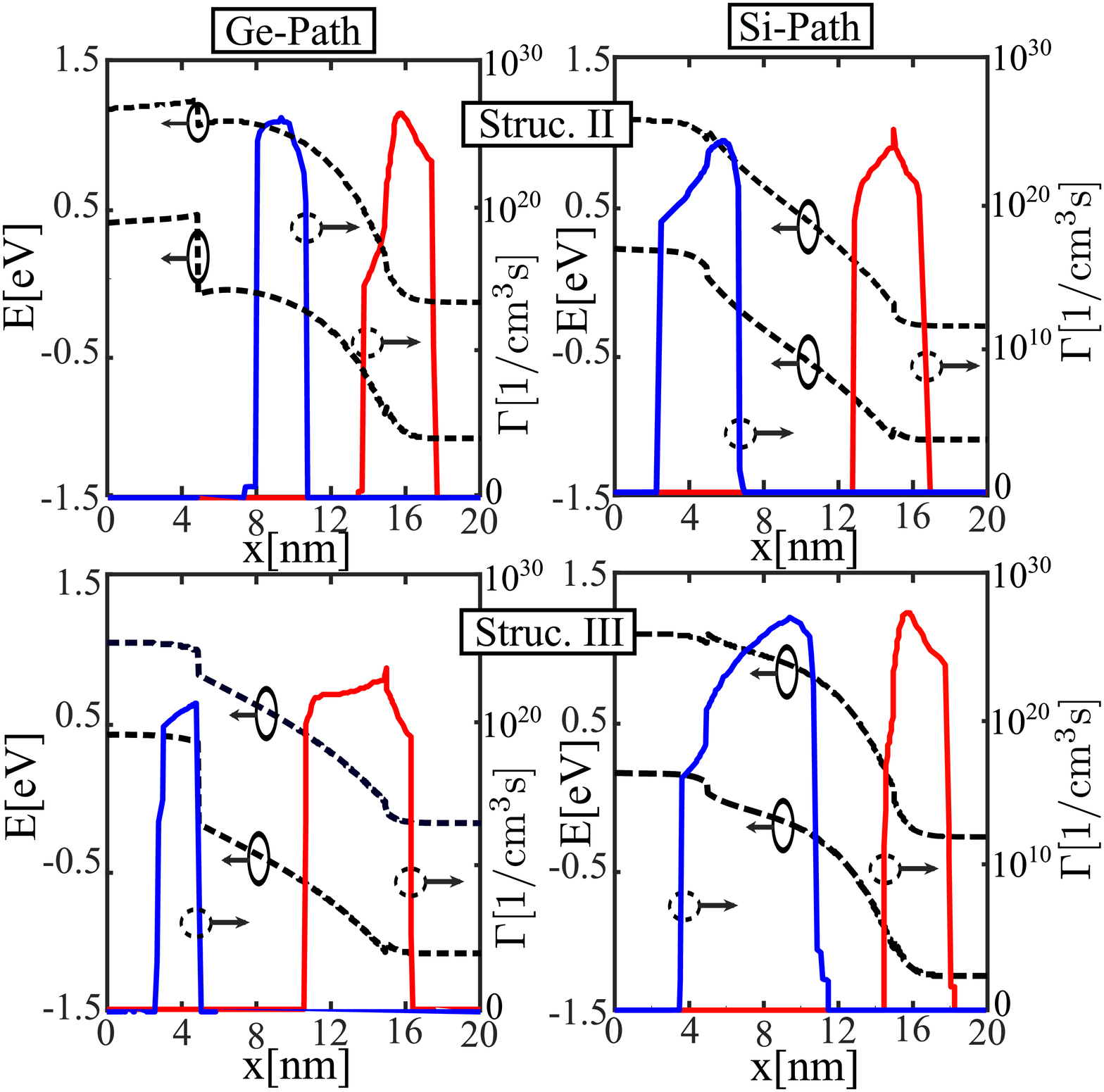}
	\caption{Energy band diagram and BTBT rate along the channel for ON-current. Figures are for structure II  at the middle of (a) Ge and (b) Si regions in the source and for structure III at the middle of (c) Si and (d) Si.}
	\label{fig:fig5}
\end{figure}

For understanding the behavior of OFF- and ON-current, energy band diagrams and BTBT generation rates ($\Gamma$) for structures II and III are plotted along the channel at OFF-state in Fig. \ref{fig:fig4}. Two paths are considered along with the device, one starts from Si in the source that we call "Si-path" and another starts from Ge and we call this one "Ge-path", see Fig. \ref{fig:fig1}. Electrons tunnel from source to channel an OFF-state. In structure II, the BTBT generation rate in the source for Ge-path is much lower than Si-path. The lower tunneling distance contributes to a higher tunneling rate. At structure III, both paths indicate the BTBT generation rate in the same range, however, the BTBT generation rate is larger for Si-path. Both regions contribute to OFF-current.

The energy band diagram and BTBT generation rate is plotted at ON-state for two structures (structure II and III) in Fig. \ref{fig:fig5}. It is obvious for structure II that one can't observe any BTBT in Ge-path. Therefore, the current totally passes through Si-path. The holes can not directly tunnel from source to channel due to long tunneling distance. Hole sees a high potential barrier from source to channel in Ge-path. This high potential barrier vanishes current flow in the Ge region, however, there is not such a barrier in Si-path.  Structure III shows a BTBT generation rate for Si-path but the BTBT generation rate in the source is larger for Ge-path. In this structure, the high potential barrier between the source and channel for Ge-path vanishes current density in the Ge region. Holes diffuse from Si at the source in the channel then tunnel from channel to source. It is obvious from the Si path in structure III that the BTBT generation rate is maximum in the channel.

\begin{figure}
	\centering
	\includegraphics[width=1.0\linewidth]{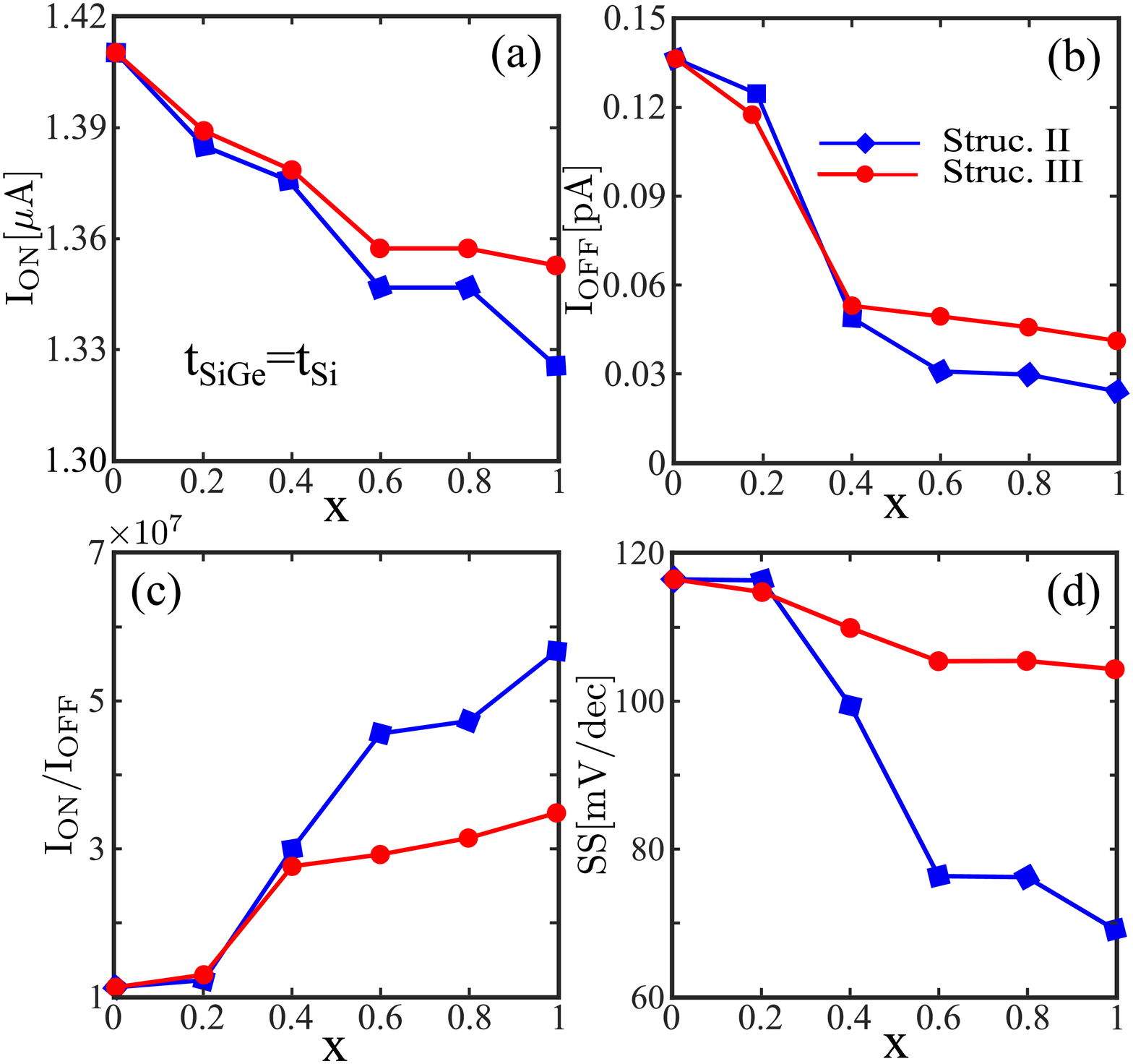}
	\caption{(a) $\mathrm{I_{ON}}$, (b) $\mathrm{I_{ON}}$, (c) ON-OFF ratio and (d) sub-threshold swing as a function of Ge mole fraction for structures II and III.}
	\label{fig:fig6}
\end{figure}

Si$_{1-x}$Ge$_{x}$ gives better compatibility with Si and is used for channel material in nowadays transistor. Here, an epi-layer SiGe is considered in the source instead of the Ge region. Ge mole fraction ($x$) is changed from $\mathrm{x=0}$ (Si) to $\mathrm{x=1}$ (Ge) and the results are plotted in Fig. \ref{fig:fig6}. ON-current reduces for both structures with increasing Ge mole fraction whereas, structure III shows more reduction. In total $\mathrm{I_{ON}}$ declines a little respect to absolute current. We observed ON-current mainly flows through the Si region so that Ge mole fraction doesn't affect ON-current. In the opposite, OFF-current decreases five times with increasing Ge mole fraction from $\mathrm{x=0}$ to $\mathrm{x=1}$. However, structure II shows a lower OFF-current. OFF-current for both structures is the same at $x$ lower than $0.4$ and for structure II decreases for $x$ larger than 0.4. ON-OFF ratio increases due to the decrease of OFF-current. The ON-OFF ratio increases five and three-time for structures II and III, respectively. $SS$ also decreases with increasing of Ge mole fraction. $\mathrm{SS}$ decreases more for structure II than structure III. $SS$ is $\mathrm{120mv/dec}$ at $\mathrm{x=0}$ (Si) whereas for $\mathrm{x=1}$  (Ge) decreases to $\mathrm{104}$ and $\mathrm{70 mV/dec}$ for structures II and III, respectively. One can observe Structure II shows a higher ON-OFF ratio and lower $\mathrm{SS}$.

\begin{figure}
	\centering
	\includegraphics[width=1.0\linewidth]{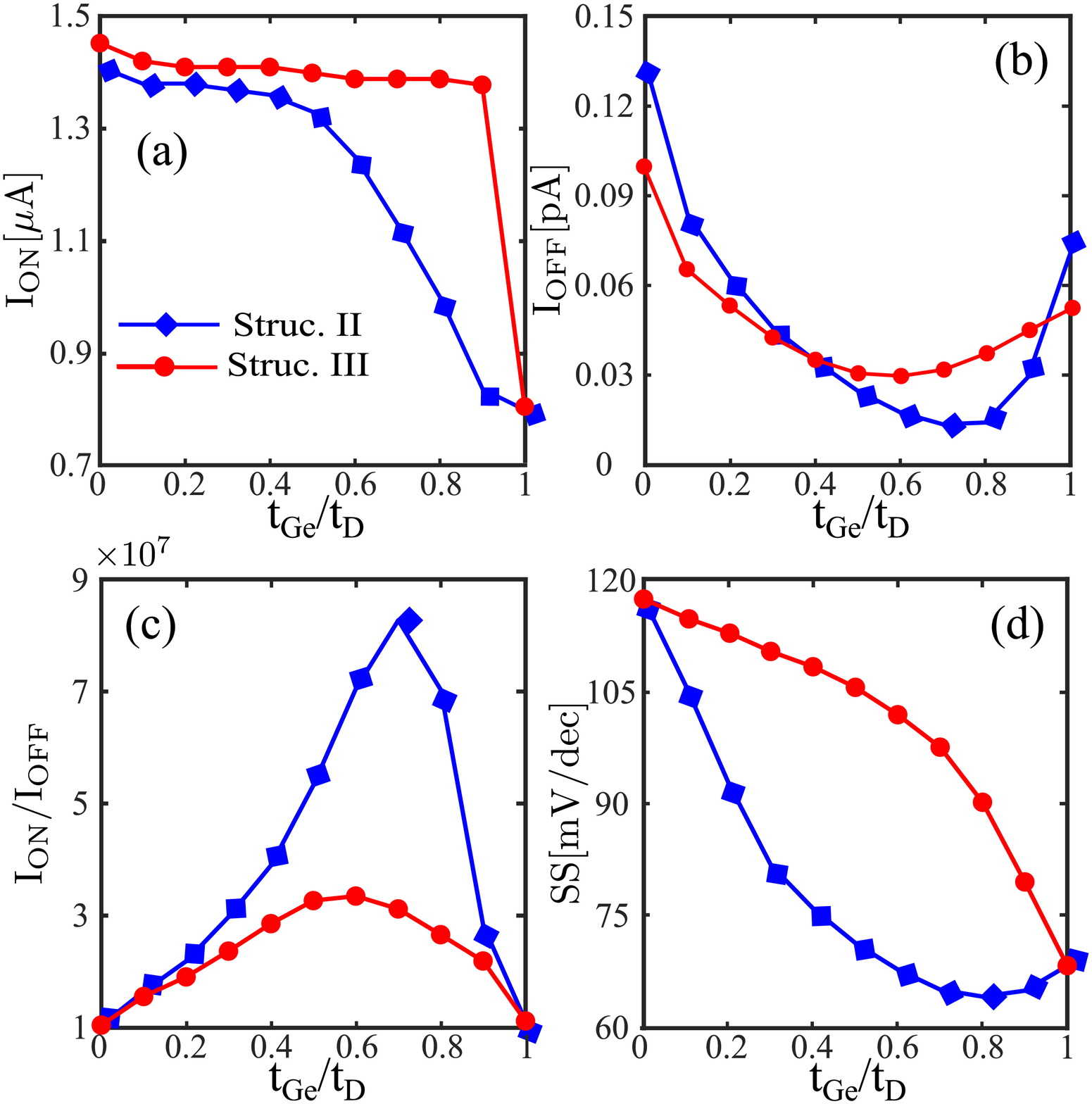}
	\caption{(a) $\mathrm{I_{ON}}$, (b) $\mathrm{I_{OFF}}$, ON-OFF ratio and sub-threshold swing as a function of Ge thickness in the source region. Ge thickness is normalized to the thickness of the source $\mathrm{L_S=L_D=5nm}$, $\mathrm{L_{Ch}=10nm}$, $\mathrm{t_{D}=5nm}$.}
	\label{fig:fig7}
\end{figure}

We observed the performance of TFET increases respect to Ge mole fraction so Ge in the source with the best performance is selected in the following. Structure II displays a better performance where the SiGe thickness is selected half of the source thickness. In the following, the effect of SiGe thickness on the performance of TFET is investigated and the results are plotted in Fig. \ref{fig:fig7}. As one can observe, $\mathrm{I_{ON}}$ decreases with increasing of Ge thickness. However, $\mathrm{I_{ON}}$ remains constant for low Ge thickness and decreases for thick Ge. In structure II, $\mathrm{I_{ON}}$ decreases for $t_{Ge}/t_D$ ($t_{Ge}$ is Ge thickness and $t_D$ is source thickness) larger than $\mathrm{0.5}$ whereas for structure III, $\mathrm{I_{ON}}$ approximately remains constant up to $\mathrm{0.9}$ then suddenly decreases. As one can see from Fig .\ref{fig:fig3}, $\mathrm{I_{ON}}$ in structure III flow near to the gate,  whereas, it passes near to substrate for structure II. When Ge thickness increase from 0 to 0.9 in structure III, $\mathrm{I_{ON}}$ remains constant because ON-current highly flows at the top of the channel and increasing of SiGe at the bottom of the source has not any considerable effect on the ON-current. But increasing Ge from 0.9 to 1, material in the top of source changes, and $\mathrm{I_{ON}}$ suddenly decreases. In structure II, the situation is inverse. ON-current passes through Si-region and increasing Ge thickness decreases Si thickness and declines ON-current gradually.
$\mathrm{I_{OFF}}$ first decreases with increasing Ge thickness then increases for high Ge thickness. Minimum $\mathrm{I_{OFF}}$ occurs at $\mathrm{t_{Ge}/t_{D}=0.6}$ and $0.8$ for structure II and III, respectively. ON-OFF ratio is reported using $\mathrm{I_{ON}}$ and $\mathrm{I_{OFF}}$. Structure II shows a higher ON-OFF ratio due to lower OFF-current. ON-OFF ratio behaves in the reverse of  $\mathrm{I_{OFF}}$, it increases for low Ge thickness and decreases for high Ge thickness. ON-OFF ratio in structure II at $\mathrm{t_{Ge}/t_{D}=0.7}$ is nine-time larger than conventional TFET and in structure III at $\mathrm{t_{Ge}/t_{D}=0.6}$ is three times larger.

Sub-threshold swing is investigated versus Ge thickness, see Fig .\ref{fig:fig7}(d). $SS$ decreases for both structures with increasing of Ge thickness whereas $SS$ for structure II increases for $\mathrm{t_{Ge}/t_{D}}$ larger than $\mathrm{0.8}$. On the other hand, structure II shows a lower $SS$ for the whole range of Ge thickness. The minimum $SS$ happens for structure II at $\mathrm{t_{Ge}/t_{D}=0.8}$ that reaches to $\mathrm{60mV/dec}$. This $SS$ is obtained for the channel with $\mathrm{L_{Ch}=10nm}$, however, lower $SS$ can be obtained for a longer channel. The best performance can be obtained for structure II where $\mathrm{t_{Ge}/t_{D}}$ is close to $\mathrm{0.7}$. In this range $\mathrm{I_{OFF}}$  and $SS$ are the minimum, the ON-OFF ratio is the maximum. $\mathrm{I_{ON}}$ only is higher for structure III and is low in this range of Ge thickness.

\section{Conclusion}
The source material of TFET is engineered to enhance performance. Three structures are compared with each other. Structure II with a SiGe layer on the top of the Si region in the source gives the best performance. This structure can decrease OFF-current four times relative to conventional TFET and increases the ON-OFF ratio three times. This structure also decreases sub-threshold swing from 117 for conventional TFET to $\mathrm{71mV/dec}$. The results showed OFF-current passes through both SiGe and Si regions, whereas, ON-current passes through the Si region with a wider bandgap. The performance of TFET increases with increasing Ge mole fraction. In the end, the best thickness for SiGe is 0.7 and 0.6 of source thickness for structures II and III, respectively.



\end{document}